\pdfoutput=1

\documentclass[12pt,a4paper]{article}

\usepackage{ifthen} 
\newboolean{pdflatex}
\setboolean{pdflatex}{true} 

\newboolean{articletitles}
\setboolean{articletitles}{true} 

\newboolean{uprightparticles}
\setboolean{uprightparticles}{true} 

\newboolean{inbibliography}
\setboolean{inbibliography}{false} 

\usepackage{microtype}
\usepackage{lineno}  
\usepackage{xspace} 
\usepackage{caption} 

\usepackage{graphicx}  
\usepackage{color}
\usepackage{colortbl}
\graphicspath{{./figs/}} 

\usepackage{amsmath} 
\usepackage{amssymb}
\usepackage{amsfonts}
\usepackage{upgreek} 

\newcommand*\patchAmsMathEnvironmentForLineno[1]{%
\expandafter\let\csname old#1\expandafter\endcsname\csname #1\endcsname
\expandafter\let\csname oldend#1\expandafter\endcsname\csname
end#1\endcsname
 \renewenvironment{#1}%
   {\linenomath\csname old#1\endcsname}%
   {\csname oldend#1\endcsname\endlinenomath}%
}
\newcommand*\patchBothAmsMathEnvironmentsForLineno[1]{%
  \patchAmsMathEnvironmentForLineno{#1}%
  \patchAmsMathEnvironmentForLineno{#1*}%
}
\AtBeginDocument{%
\patchBothAmsMathEnvironmentsForLineno{equation}%
\patchBothAmsMathEnvironmentsForLineno{align}%
\patchBothAmsMathEnvironmentsForLineno{flalign}%
\patchBothAmsMathEnvironmentsForLineno{alignat}%
\patchBothAmsMathEnvironmentsForLineno{gather}%
\patchBothAmsMathEnvironmentsForLineno{multline}%
\patchBothAmsMathEnvironmentsForLineno{eqnarray}%
}

\usepackage{hyperref}    
\usepackage[all]{hypcap} 

\def\lhcb {\mbox{LHCb}\xspace}

\def\MagUp {\mbox{\em Mag\kern -0.05em Up}\xspace}

\ifthenelse{\boolean{uprightparticles}}%
{

 \def\Pmu         {\ensuremath{\upmu}\xspace}

 \def\Ptau        {\ensuremath{\uptau}\xspace}

 \def\PDelta      {\ensuremath{\Delta}\xspace}                 
 \def\PXi      {\ensuremath{\Xi}\xspace}                 
 \def\PLambda      {\ensuremath{\Lambda}\xspace}                 
 \def\PSigma      {\ensuremath{\Sigma}\xspace}                 
 \def\POmega      {\ensuremath{\Omega}\xspace}                 
 \def\PUpsilon      {\ensuremath{\Upsilon}\xspace}                 
 
 \def\PB      {\ensuremath{\mathrm{B}}\xspace}                 
                  
 \def\PD      {\ensuremath{\mathrm{D}}\xspace}

 \def\PK      {\ensuremath{\mathrm{K}}\xspace}

 \def\PW      {\ensuremath{\mathrm{W}}\xspace}

 \def\PZ      {\ensuremath{\mathrm{Z}}\xspace}                 
                  
 \def\Pb      {\ensuremath{\mathrm{b}}\xspace}                 
 \def\Pc      {\ensuremath{\mathrm{c}}\xspace}                 
                  
 \def\Pe      {\ensuremath{\mathrm{e}}\xspace}

 \def\Pi      {\ensuremath{\mathrm{i}}\xspace}

 \def\Pp      {\ensuremath{\mathrm{p}}\xspace}

 \def\Pt      {\ensuremath{\mathrm{t}}\xspace}

}
{

 \def\Pmu         {\ensuremath{\mu}\xspace}

 \def\Ptau        {\ensuremath{\tau}\xspace}

 \mathchardef\PDelta="7101
 \mathchardef\PXi="7104
 \mathchardef\PLambda="7103
 \mathchardef\PSigma="7106
 \mathchardef\POmega="710A
 \mathchardef\PUpsilon="7107
                  
 \def\PB      {\ensuremath{B}\xspace}                 
                  
 \def\PD      {\ensuremath{D}\xspace}

 \def\PK      {\ensuremath{K}\xspace}

 \def\PW      {\ensuremath{W}\xspace}

 \def\PZ      {\ensuremath{Z}\xspace}                 
                  
 \def\Pb      {\ensuremath{b}\xspace}                 
 \def\Pc      {\ensuremath{c}\xspace}                 
                  
 \def\Pe      {\ensuremath{e}\xspace}

 \def\Pi      {\ensuremath{i}\xspace}

 \def\Pp      {\ensuremath{p}\xspace}

 \def\Pt      {\ensuremath{t}\xspace}

}

\makeatletter
\ifcase \@ptsize \relax
  \newcommand{\miniscule}{\@setfontsize\miniscule{4}{5}}
\or
  \newcommand{\miniscule}{\@setfontsize\miniscule{5}{6}}
\or
  \newcommand{\miniscule}{\@setfontsize\miniscule{5}{6}}
\fi
\makeatother

\DeclareRobustCommand{\optbar}[1]{\shortstack{{\miniscule (\rule[.5ex]{1.25em}{.18mm})}
  \\ [-.7ex] $#1$}}

\def\en         {{\ensuremath{\Pe^-}}\xspace}   
\def\ep         {{\ensuremath{\Pe^+}}\xspace}
\def\epm        {{\ensuremath{\Pe^\pm}}\xspace} 
\def\epem       {{\ensuremath{\Pe^+\Pe^-}}\xspace}

\def\mumu       {{\ensuremath{\Pmu^+\Pmu^-}}\xspace}

\def\tautau     {{\ensuremath{\Ptau^+\Ptau^-}}\xspace}

\def\Wp     {{\ensuremath{\PW^+}}\xspace}
\def\Wm     {{\ensuremath{\PW^-}}\xspace}
\def\Wpm    {{\ensuremath{\PW^\pm}}\xspace}
\def\Z      {{\ensuremath{\PZ}}\xspace}

\def\cquark    {{\ensuremath{\Pc}}\xspace}
\def\cquarkbar {{\ensuremath{\overline \cquark}}\xspace}
\def\ccbar     {{\ensuremath{\cquark\cquarkbar}}\xspace}
\def\bquark    {{\ensuremath{\Pb}}\xspace}
\def\bquarkbar {{\ensuremath{\overline \bquark}}\xspace}
\def\bbbar     {{\ensuremath{\bquark\bquarkbar}}\xspace}
\def\tquark    {{\ensuremath{\Pt}}\xspace}
\def\tquarkbar {{\ensuremath{\overline \tquark}}\xspace}
\def\ttbar     {{\ensuremath{\tquark\tquarkbar}}\xspace}

  \def\Kbar    {{\kern 0.2em\overline{\kern -0.2em \PK}{}}\xspace}

\def\KorKbar    {\kern 0.18em\optbar{\kern -0.18em K}{}\xspace}

  \def\Dbar    {{\kern 0.2em\overline{\kern -0.2em \PD}{}}\xspace}

\def\DorDbar    {\kern 0.18em\optbar{\kern -0.18em D}{}\xspace}

\def\Bbar    {{\ensuremath{\kern 0.18em\overline{\kern -0.18em \PB}{}}}\xspace}

\def\BorBbar    {\kern 0.18em\optbar{\kern -0.18em B}{}\xspace}

  \def\Y#1S{\ensuremath{\PUpsilon{(#1S)}}\xspace}

\def\proton      {{\ensuremath{\Pp}}\xspace}

\def\Lbar        {{\ensuremath{\kern 0.1em\overline{\kern -0.1em\PLambda}}}\xspace}
\def\LorLbar    {\kern 0.18em\optbar{\kern -0.18em \PLambda}{}\xspace}

\def\to                 {\ensuremath{\rightarrow}\xspace}

\def\order   {{\ensuremath{\mathcal{O}}}\xspace}

\newcommand{\as}{{\ensuremath{\alpha_s}}\xspace}

\def\AT#1     {\ensuremath{A_{\mathrm{T}}^{#1}}\xspace}           

\def\C#1      {\ensuremath{\mathcal{C}_{#1}}\xspace}                       
\def\Cp#1     {\ensuremath{\mathcal{C}_{#1}^{'}}\xspace}                    
\def\Ceff#1   {\ensuremath{\mathcal{C}_{#1}^{\mathrm{(eff)}}}\xspace}        
\def\Cpeff#1  {\ensuremath{\mathcal{C}_{#1}^{'\mathrm{(eff)}}}\xspace}       
\def\Ope#1    {\ensuremath{\mathcal{O}_{#1}}\xspace}                       
\def\Opep#1   {\ensuremath{\mathcal{O}_{#1}^{'}}\xspace}                    

\newcommand{\tev}{\ifthenelse{\boolean{inbibliography}}{\ensuremath{~T\kern -0.05em eV}\xspace}{\ensuremath{\mathrm{\,Te\kern -0.1em V}}}\xspace}
\newcommand{\gev}{\ensuremath{\mathrm{\,Ge\kern -0.1em V}}\xspace}
\newcommand{\mev}{\ensuremath{\mathrm{\,Me\kern -0.1em V}}\xspace}
\newcommand{\kev}{\ensuremath{\mathrm{\,ke\kern -0.1em V}}\xspace}
\newcommand{\ev}{\ensuremath{\mathrm{\,e\kern -0.1em V}}\xspace}
\newcommand{\gevc}{\ensuremath{{\mathrm{\,Ge\kern -0.1em V\!/}c}}\xspace}
\newcommand{\mevc}{\ensuremath{{\mathrm{\,Me\kern -0.1em V\!/}c}}\xspace}
\newcommand{\gevcc}{\ensuremath{{\mathrm{\,Ge\kern -0.1em V\!/}c^2}}\xspace}
\newcommand{\gevgevcccc}{\ensuremath{{\mathrm{\,Ge\kern -0.1em V^2\!/}c^4}}\xspace}
\newcommand{\mevcc}{\ensuremath{{\mathrm{\,Me\kern -0.1em V\!/}c^2}}\xspace}

\def\mum  {\ensuremath{{\,\upmu\rm m}}\xspace}

\def\invpb {\ensuremath{\mbox{\,pb}^{-1}}\xspace}

\def\invfb   {\ensuremath{\mbox{\,fb}^{-1}}\xspace}

\def\order{{\ensuremath{\cal O}}\xspace}
\newcommand{\chisq}{\ensuremath{\chi^2}\xspace}

\def\gsim{{~\raise.15em\hbox{$>$}\kern-.85em
          \lower.35em\hbox{$\sim$}~}\xspace}
\def\lsim{{~\raise.15em\hbox{$<$}\kern-.85em
          \lower.35em\hbox{$\sim$}~}\xspace}

\def\sqs   {\ensuremath{\protect\sqrt{s}}\xspace}

\def\ptot       {\mbox{$p$}\xspace}
\def\pt         {\mbox{$p_{\rm T}$}\xspace}

\newcommand{\lum} {\ensuremath{\mathcal{L}}\xspace}

\def\fewz       {\mbox{\textsc{Fewz}}\xspace}

\def\geant      {\mbox{\textsc{Geant4}}\xspace}

\def\powheg     {\mbox{\textsc{Powheg}}\xspace}
\def\pythia     {\mbox{\textsc{Pythia}}\xspace}
\def\resbos     {\mbox{\textsc{ResBos}}\xspace}

\def\tell1  {TELL1\xspace}
\def\ukl1   {UKL1\xspace}

\newcommand{\ie}{\mbox{\itshape i.e.}\xspace}

\usepackage{cite} 
\usepackage{mciteplus}

\usepackage{longtable} 

\usepackage{rotating}

\begin{document}

\renewcommand{\thefootnote}{\fnsymbol{footnote}}
\setcounter{footnote}{1}


\begin{titlepage}
\pagenumbering{roman}

\vspace*{-1.5cm}
\centerline{\large EUROPEAN ORGANIZATION FOR NUCLEAR RESEARCH (CERN)}
\vspace*{1.5cm}
\hspace*{-0.5cm}
\begin{tabular*}{\linewidth}{lc@{\extracolsep{\fill}}r}
\ifthenelse{\boolean{pdflatex}}
{\vspace*{-2.7cm}\mbox{\!\!\!\includegraphics[width=.14\textwidth]{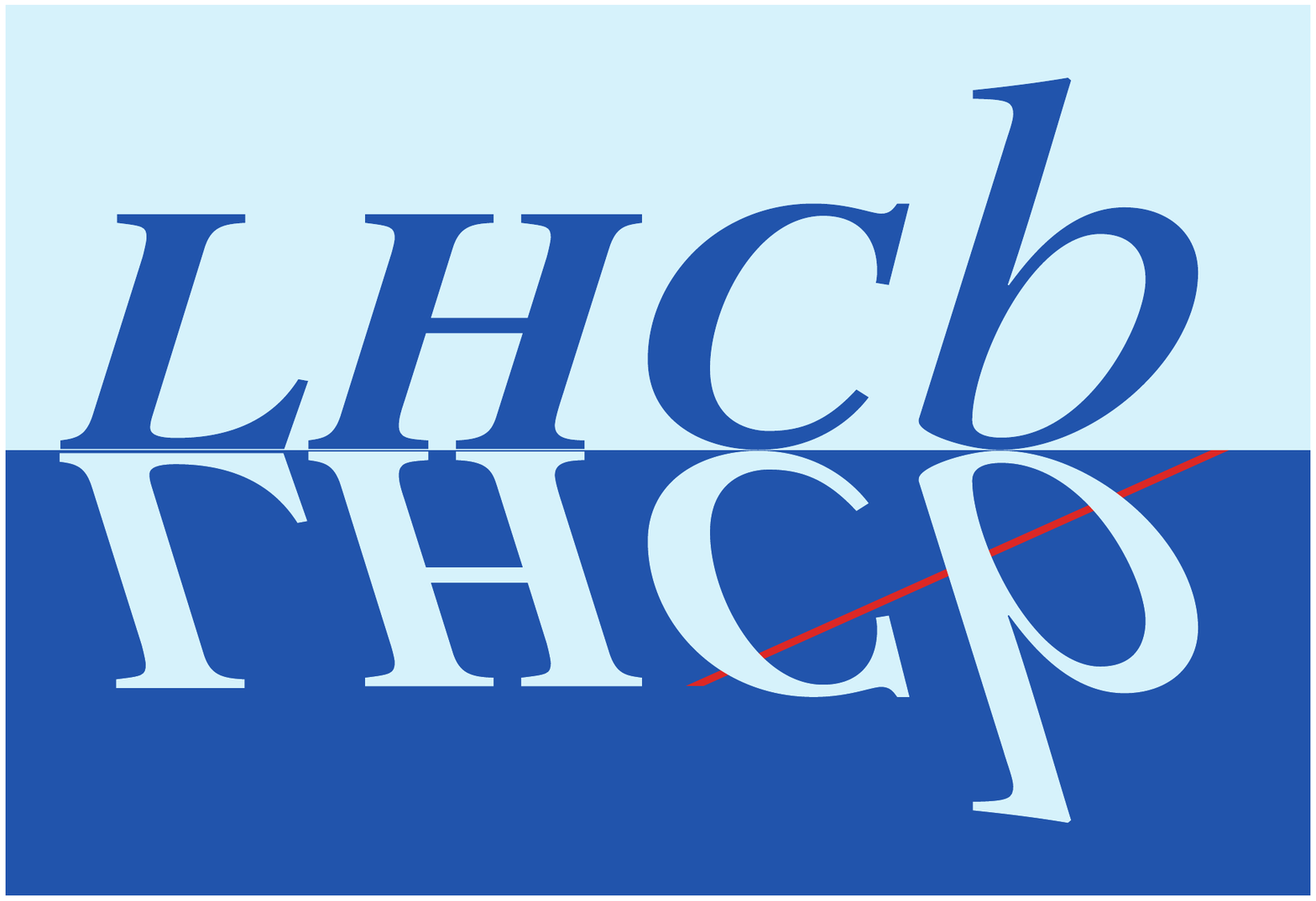}} & &}%
{\vspace*{-1.2cm}\mbox{\!\!\!\includegraphics[width=.12\textwidth]{lhcb-logo.eps}} & &}%
\\
 & & CERN-PH-EP-2015-045 \\  
 & & LHCb-PAPER-2015-003 \\  
 & & 3 March 2015 \\ 
\end{tabular*}

\vspace*{2.0cm}

{\bf\boldmath\huge
\begin{center}
    Measurement of forward $\Z\to\epem$ production at $\sqs=8$\tev
\end{center}
}

\vspace*{1.0cm}

\begin{center}
The LHCb collaboration\footnote{Authors are listed at the end of this paper.}
\end{center}

\vspace{\fill}

\begin{abstract}
  \noindent
A measurement of the cross-section for $\Z$-boson production in the forward region 
of \proton\proton collisions
at 8\tev centre-of-mass energy is presented.  The measurement is based on a sample 
of $\Z\to\epem$ decays reconstructed
using the \lhcb\ detector, corresponding 
to an integrated luminosity of 2.0\invfb.  
The acceptance is
defined by the requirements $2.0<\eta<4.5$ and $\pt>20$\gev for the pseudorapidities and transverse 
momenta of the leptons.  Their invariant mass is required to lie in the range 
60--120\gev.  The cross-section is determined to be
$$ \sigma(\proton\proton\to\Z\to\epem)=93.81\pm0.41({\rm stat})\pm1.48({\rm syst})\pm1.14({\rm lumi})\;{\rm pb}\,,$$
where the first uncertainty is statistical and the second reflects all systematic effects apart from that arising 
from the luminosity, which is given as the third uncertainty.  Differential cross-sections are presented as functions of the
Z-boson rapidity and of the angular variable $\phi^*$, which is related to the
$\Z$-boson transverse momentum.  
\end{abstract}

\vspace*{0.5cm}

\begin{center}
  Submitted to JHEP 
\end{center}

\vspace{\fill}

{\footnotesize 
\centerline{\copyright~CERN on behalf of the \lhcb collaboration, license \href{http://creativecommons.org/licenses/by/4.0/}{CC-BY-4.0}.}}
\vspace*{1.5mm}

\end{titlepage}


\newpage
\setcounter{page}{2}
\mbox{~}
%
%
%
%

\cleardoublepage


\renewcommand{\thefootnote}{\arabic{footnote}}
\setcounter{footnote}{0}



\pagestyle{plain} 
\setcounter{page}{1}
\pagenumbering{arabic}


%


 

\newpage

\section{Introduction}
\label{sec:Introduction}

Measurements of vector boson production in hadron collisions 
permit tests of electroweak physics and of QCD. 
The effective kinematic range of the \lhcb detector\cite{Alves:2008zz}, 
approximately $2.0<\eta<4.5$ where $\eta$ is the pseudorapidity, 
complements that of the general purpose LHC detectors 
whose acceptance for precise measurements extends to $|\eta|\approx2.5$.
Measurements at \lhcb\ are sensitive to the knowledge of the proton structure 
functions at very low Bjorken $x$ values, where  the parton distribution functions (PDFs) 
are not well constrained by previous data, 
or by other LHC experiments~\cite{Thorne:2008am}.

The most straightforward decay mode in which \lhcb\ can study the production of the \Z\ 
boson is the channel
$\Z\to\mumu$\cite{LHCb-PAPER-2012-008}, since the experiment has a highly efficient trigger 
and precise reconstruction capabilities for high-momentum muons.  
However, it is also desirable to examine the channel
$\Z\to\epem$,\footnote{Throughout this note we use 
$\Z\to\epem$ to imply the formation of \epem\ through either a \Z or a virtual 
photon $\gamma^*$, including the effect of their interference.}
which offers a statistically independent sample, with
substantially different sources of systematic uncertainties.   
The main difficulty with electron reconstruction in \lhcb\ is the energy measurement.  
A significant amount of material is traversed by the particles before they reach the 
magnet, and their measured momenta are therefore 
frequently reduced by bremsstrahlung, which cannot 
be recovered fully using the calorimeters because 
of saturation at an energy corresponding to 
a transverse momentum of around 10\gev per channel. 
Consequently the initial electron directions are well determined, 
but their measured momenta are low by a variable amount, $\sim25\%$ on average.
Therefore the rapidity of the \Z\ boson, $y_{\Z}$, is well determined   
while its transverse momentum, \pt, is poorly measured and biased. 
Thus, as in Ref.~\cite{LHCb-PAPER-2012-036}, 
the distribution of the angular variable $\phi^*$, which is correlated with \pt 
but less affected by bremsstrahlung, is studied. 
The $\phi^*$ variable~\cite{Banfi:2010cf} is defined in terms of the directions of the particle momenta
\[ \phi^*\equiv\frac{\tan(\phi_{\mathrm{acop}}/2)}{\cosh(\Delta\eta/2)}\approx\frac{\pt}{M}\,, \]
where $\Delta\eta$ is the difference in pseudorapidity between the leptons, 
the acoplanarity angle $\phi_{\mathrm{acop}}=\pi-|\Delta\phi|$ depends on the difference 
between the azimuthal directions of the lepton momenta, $\Delta\phi$, 
and $M$ and $\pt$ are the invariant mass and transverse momentum 
of the lepton pair.\footnote{Natural units with $\hbar=c=1$ 
are used throughout, so that mass and momentum are measured in units of energy.} 

In this paper, we present a measurement of the cross-section for $\proton\proton\to\Z\to\epem$ using the data
recorded by \lhcb\ in 2012 at $\sqs=8$\tev, corresponding to an integrated luminosity of 2.0\invfb.  
The approach is essentially the same as that used in a
previous study of the same channel at $\sqs=7$\tev~\cite{LHCb-PAPER-2012-036}.
The current measurement, as well as being at a higher centre-of-mass energy, benefits from a significantly 
higher and more precisely determined integrated luminosity, and from stable trigger conditions. 
Futhermore, an improved Monte Carlo simulation of the detector and an improved 
modelling of electron bremsstrahlung are available. 
The determination is performed in the same kinematic region 
as the \lhcb\ measurement of $\Z\to\mumu$~\cite{LHCb-PAPER-2012-008}, namely $60<M(\Z)<120$\gev and 
$2.0<\eta<4.5$ and $\pt>20\gev$ for the leptons.

Section~\ref{sec:Detector} gives a brief description of the detector, triggers and software, after which 
Sect.~\ref{sec:Selection} describes the event selection and the analysis procedure, and 
Sect.~\ref{sec:systematics} explains the techniques used for determining the main uncertainties in the measurement.  
The results are presented in Sect.~\ref{sec:Results} followed by a brief summary in Sect.~\ref{sec:Summary}.

\section{Detector and simulation}
\label{sec:Detector}
The \lhcb detector~\cite{Alves:2008zz,LHCb-DP-2014-002} is a single-arm forward
spectrometer designed for the study of particles containing \bquark or \cquark
quarks. The detector includes a high-precision tracking system
consisting of a silicon-strip vertex detector surrounding the pp
interaction region,
a large-area silicon-strip detector located
upstream of a dipole magnet with a bending power of about
$4{\rm\,Tm}$, and three stations of silicon-strip detectors and straw
drift tubes, 
placed downstream of the magnet.
The tracking system provides a measurement of momentum, \ptot,  with
a relative uncertainty that varies from 0.5\% at low momentum to 1.0\% at 200\gev.
The minimum distance of a track to a primary vertex, the impact parameter, is measured with a resolution 
of $(15+29/\pt)\mum$, where \pt is in\,\gev.
Different types of charged hadrons are distinguished using information
from two ring-imaging Cherenkov detectors.
Photon, electron and
hadron candidates are identified by a calorimeter system consisting of
scintillating-pad (SPD) and preshower (PRS) detectors, an electromagnetic
calorimeter (ECAL) and a hadronic calorimeter (HCAL). Muons are identified by a
system composed of alternating layers of iron and multiwire
proportional chambers. 

The trigger 
consists of a hardware stage, based on information from the calorimeter and muon
systems, followed by a software stage, which applies a full event
reconstruction.
Events are first required to pass the hardware trigger for electrons, which selects events having 
an electromagnetic cluster of high transverse energy geometrically associated with signals in the
PRS and SPD detectors. 
This high-\pt\ single-electron trigger is then refined by the software trigger.   
Global event cuts on the numbers of hits in several detectors, such as the SPD, 
are applied in order to prevent high-multiplicity events from dominating the processing time. 

Simulated event samples of $\Z\to\epem$ with $M(\epem)>40$\gev are used in the analysis.
Simulated samples of $\Z\to\tautau$, $\ttbar$, $\Wp\Wm$ and $\Wpm\Z$  
are used to assess possible background contributions.  
In the simulation, pp collisions are generated using
\pythia~8.1~\cite{Sjostrand:2006za,*Sjostrand:2007gs}  with a specific \lhcb
configuration~\cite{LHCb-PROC-2010-056}. 
The interaction of the generated particles with the detector and its
response are implemented using the \geant
toolkit~\cite{Allison:2006ve, *Agostinelli:2002hh} as described in
Ref.~\cite{LHCb-PROC-2011-006}.


\section{Data analysis}
\label{sec:Selection}

The reconstructed particles used as the basis of the analysis 
satisfy basic track quality requirements.
The following criteria are applied in order to refine the sample of candidates for analysis:
\begin{itemize}
 \item either the electron or the positron candidate should satisfy a single-electron trigger
at all stages of the trigger;
 \item the reconstructed electron and positron candidates should have pseudorapidity and transverse 
  momentum satisfying $2.0<\eta<4.5$ and  
 $\pt>20$\gev respectively.  The reconstructed \epem invariant mass should be greater than 40\gev; 
 \item to ensure good track quality, 
the electron and positron candidates should have momenta measured with estimated 
fractional uncertainty smaller than $10\%$;  
\item in order to identify the particles as electron candidates,
both are required to show associated 
energy deposition in the calorimeters characteristic of high-energy electrons, namely
  $E_{\rm ECAL}/p>0.1$, $E_{\rm HCAL}/p<0.05$ and $E_{\rm PRS}>50$\mev where $E_{\rm ECAL}$, $E_{\rm HCAL}$ and
$E_{\rm PRS}$ denote the energies recorded in the electromagnetic calorimeter, 
hadronic calorimeter and preshower detector respectively;
 \item if more than one $\epem$ pair satisfies the above criteria in an event
(which occurs in approximately 0.7\% of cases), one is selected at random.
\end{itemize} 
Applying these selection requirements on data, 65\,552 $\Z\to\epem$ candidates are obtained.  
Most backgrounds are removed by subtracting a sample of same-sign $\epm\epm$ candidates. 
The validity of this procedure is assessed in Sect.~\ref{sec:systematics}.
Applying identical selection criteria, 4595 same-sign candidates are found. 

The cross-section is determined using the following expression:
\begin{equation}
 \sigma = \frac{N(\epem)-N(\epm\epm) -N_{\rm bg}}
 {\epsilon\cdot
   \int\lum\mathrm{d}t}
\cdot f_{\mathrm{MZ}}\,,
\label{equ:Corr}
\end{equation}
where $N(\epem)$ is the number of signal candidates selected in data, $N(\epm\epm)$ 
is the number of same-sign candidates,
$N_{\rm bg}$ is the expected background not covered by same-sign candidates 
(predominantly $\Z\to\tautau$) taken from simulation 
and $\int\lum\mathrm{d}t$ is the integrated luminosity.  
The event detection efficiency, $\epsilon$, is evaluated using a combination of data and 
simulation as explained below, and refers to events 
for which the true electrons satisfy $2.0<\eta<4.5$, $\pt>20\gev$ and $60<M(\epem)<120$\gev.
The factor $f_{\mathrm{MZ}}$ (determined from simulation) corrects for the inclusion in the data sample of
$\Z\rightarrow\epem$ 
candidates that pass the event selection even though their true mass lies outside the range $60<M(\epem)<120$\gev. 
The correction procedure is applied for 17 bins in \Z\ rapidity in the range 
$2.00<y_{\Z}<4.25$.\footnote{Although tracks can be reconstructed at larger rapidities,
the efficiency for electron identification vanishes just above $y_{\Z}=4.25$ because of 
the inner edge of the calorimeter acceptance, 
so that no candidate event at higher rapidity survives in either data or simulation.}  The analysis is also performed for 15 bins in $\phi^*$.
The choice of binning takes account of the available sample size and the resolutions achieved on
$y_{\Z}$ and $\phi^*$.  
The luminosity is obtained with an overall uncertainty of $1.2\%$\cite{Aaij:2014ida}.  

The efficiency of the event selection is factorised into several components,
\begin{equation}
 \epsilon=
  \epsilon_{\mathrm{track}}\cdot
  \epsilon_{\mathrm{kin}}\cdot
  \epsilon_{\mathrm{PID}}\cdot
  \epsilon_{\mathrm{GEC}}\cdot
  \epsilon_{\mathrm{trig}}\,.
\end{equation}
The efficiencies are determined such that the efficiency for each stage of the analysis is 
estimated for events that pass the preceding stages.
Thus,
$\epsilon_{\mathrm{track}}$ is the efficiency associated with the reconstruction of both electrons 
as tracks satisfying the quality requirements and 
$\epsilon_{\mathrm{kin}}$ gives the efficiency that both these
reconstructed electron tracks satisfy the kinematic 
acceptance requirements on $\pt$ and $\eta$.
Similarly, $\epsilon_{\mathrm{PID}}$ is the efficiency for identification of the tracks as electrons,
$\epsilon_{\mathrm{GEC}}$ is the estimated efficiency of 
the global event cuts for these events and
$\epsilon_{\mathrm{trig}}$ is the trigger efficiency. 
The determination of these contributions 
to the efficiency is performed separately 
in each bin of $y_{\Z}$ or of $\phi^*$. 
The contributions that the terms in the efficiency make to the systematic 
uncertainty on the measurement are summarised in Sect.~\ref{sec:systematics}.

The tracking efficiency, $\epsilon_{\rm track}$, gives the probability that, in events in which the 
electrons satisfy 
$2.0<\eta<4.5$, $\pt>20$\gev and $60<M(\epem)<120$\gev at the particle level
(\ie\ based on their true momenta, as known in simulated events),  
both of them correspond to reconstructed tracks satisfying the track quality requirements.
In order to characterise accurately the dependence of the efficiency 
on $y_{\Z}$ and $\phi^*$, the efficiency is taken from 
simulation. A consistency check using data allows 
a systematic uncertainty to be assigned as described in Sect.~\ref{sec:systematics}.
The efficiency shows a significant dependence on $y_{\rm Z}$, 
but almost no variation with $\phi^*$.   
 
The kinematic efficiency, $\epsilon_{\rm kin}$, 
accounts for the possibility that a $\Z\to\epem$ decay in which  
the electron momenta at particle level satisfy the kinematic requirements on \pt\ and $\eta$ may not do so for 
the reconstructed momenta, even though the tracks are reconstructed and satisfy the track quality requirements.
This is a sizeable correction because of bremsstrahlung.  The efficiency is determined 
using simulated events, with data used to assess the systematic uncertainty as described in Sect.~\ref{sec:systematics}. 
In contrast to the earlier analysis~\cite{LHCb-PAPER-2012-036}, 
the particle-level electron momentum used here is the momentum before final-state radiation (FSR), so that the 
kinematic efficiency also includes the effect of FSR as implemented in \pythia 8.1, and an additional correction 
is no longer applied.    

The particle identification efficiency, $\epsilon_{\rm PID}$, 
accounts for the possibility that a signal event passing 
all track finding and kinematic requirements fails the electron identification requirements, 
either because the track falls 
outside the calorimeter acceptance, or because the requirements on calorimeter energies 
are not satisfied.
The overall efficiency is taken from simulation in order to model accurately a significant dependence on $y_{\Z}$.
This dependence is a consequence of the geometrical acceptance and is assumed to be modelled reliably.
The efficiency of the calorimeter energy requirements is validated using data as described in Sect.~\ref{sec:systematics}, 
leading to a systematic uncertainty. 

The only global event cut that has any significant effect on the $\Z\to\epem$ channel is a requirement 
in the electron triggers that
the number of hits in the SPD be less than 600.  In order to assess the consequent 
loss of events, use is made of  
$\Z\to\mumu$ events that satisfy the dimuon trigger, for which the number of SPD hits, $N_{\rm SPD}$, 
is required to be less than 900.  The distribution of $N_{\rm SPD}$ in dielectron events should be the
same as that in dimuon events apart from the contribution from the leptons.  In the region $N_{\rm SPD}<600$  
it is observed that the distribution for $\Z\to\epem$ is consistent with that for $\Z\to\mumu$
with an upward shift of $10\pm5$ hits associated with showering of the electron and positron.
This shift is confirmed in simulation.  
Accordingly, we use the fraction of $\Z\to\mumu$ dimuon triggers in the range
$590\le N_{\rm SPD}<900$ to estimate the loss of dielectron events in the range  $600\le N_{\rm SPD}<910$.
The small fraction of dimuon events having $ N_{\rm SPD}\ge 900$ is estimated to be 0.7\% by extrapolation 
using an empirical fit to the distribution.   
The efficiency shows a weak dependence on \Z\ rapidity and $\phi^*$.

The trigger efficiencies for events passing the selection cuts 
are determined from data using a ``tag-and-probe'' method. 
The principle is to use events in which the electron satisfies the trigger  
to determine the efficiency for the positron to satisfy the trigger and vice versa.  

\begin{figure}[tb]
  \begin{center}
    \includegraphics[scale=0.38]{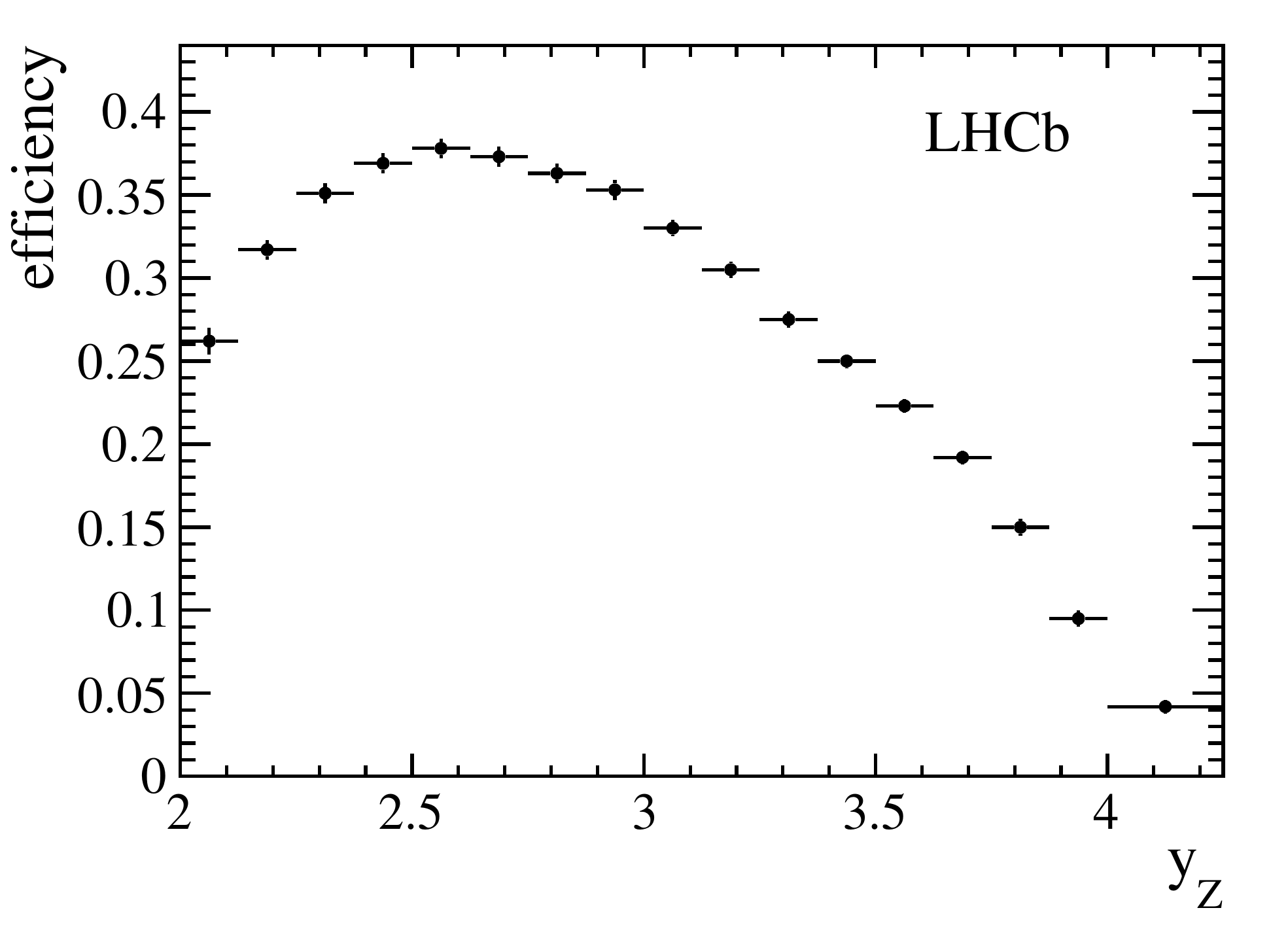}
    \includegraphics[scale=0.38]{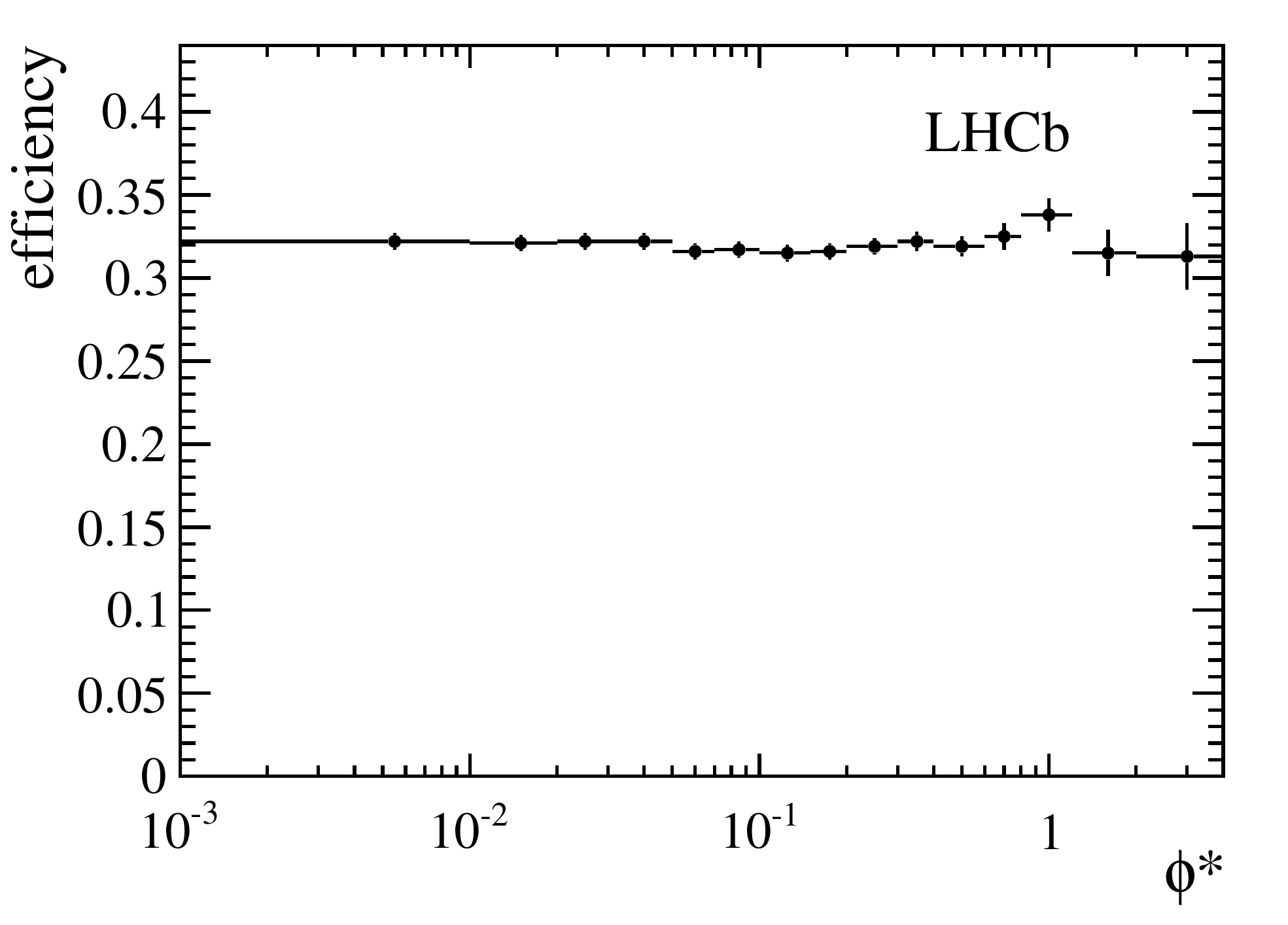}
    \vspace*{-1.0cm}
  \end{center}
  \caption{Overall detection efficiency, $\epsilon$, 
determined from a combination of data and simulation as described in the text,  
shown as a function of (left) $y_{\Z}$ and (right) $\phi^*$.}
  \label{fig:effOverall}
\end{figure}

The values of the various correction factors, 
averaged over all values of $y_{\Z}$, are summarised in Table~\ref{tab:effic}.
The uncorrelated components of the uncertainties are generally related to the statistical uncertainty 
associated with each efficiency.  These can be sizeable in individual bins, 
but become small when their effect on the integrated cross-section is considered. 
The correlated components are taken to be fully correlated between bins, and 
therefore have roughly the same effect on individual bins and on
the integrated cross-section.  The principal systematic uncertainties are discussed 
in greater detail in Sect.~\ref{sec:systematics}.
The overall efficiency is shown as a function of $y_{\Z}$ and $\phi^*$ in 
Fig.~\ref{fig:effOverall}.

\begin{table}[tbh]
\caption{Efficiencies and other factors used for the cross-section determination (see Eq.~(\ref{equ:Corr}))
averaged over the experimental acceptance by integrating over $y_{\Z}$. 
The fractional uncertainties on the overall factors are also given, 
separated into components that are assumed to be correlated and uncorrelated 
between bins of the differential distributions.}
\begin{center}\begin{tabular}{lccc}
\hline
                              &  & \multicolumn{2}{c}{Fractional uncertainty}  \\ 
                              & Average value & Uncorrelated & Correlated \\ 
  \hline
$\epsilon_{\mathrm{track}}$ &  0.912 & 0.001 & 0.010  \\
$\epsilon_{\mathrm{kin}}$ & 0.507 & 0.002 & 0.006  \\
$\epsilon_{\mathrm{PID}}$  &  0.838 & 0.001 & 0.007 \\
$\epsilon_{\mathrm{GEC}}$ & 0.916 &  ---  & 0.006 \\
$\epsilon_{\mathrm{trig}}$ & 0.892 & 0.001 & --- \\
\hline
$\epsilon$ & 0.319& 0.002 & 0.016  \\
$f_{\mathrm{MZ}}$   &  0.969 & 0.001 &  ---\\ 
Background estimation & ---& ---& 0.004 \\
$\int\lum{\rm d}t$ /\invpb & 1976& --- & 0.012 \\
\hline
\end{tabular}\end{center}
\label{tab:effic}
\end{table}


\section{Systematic uncertainties}
\label{sec:systematics}

The tracking efficiency is evaluated using simulation and checked using 
data.  The principle is to search for events where a track appears to be missing, 
so that the signature of a \Z becomes a high-\pt\ electron track accompanied by a 
high-energy ECAL cluster with no associated track. An efficiency is determined 
by comparison with the corresponding number
of events in which two tracks are reconstructed.  
To reduce background, more stringent particle identification requirements are imposed
on the tag electron, and isolation requirements are imposed on both the electron and the ECAL cluster.
This cannot be regarded as a direct measurement of the efficiency of 
the main analysis selection because more stringent requirements are employed.  
Instead, the same procedure is applied
to the simulated event sample.  The ratio between the two, $0.990\pm0.004$, 
is taken as a correction to $\epsilon_{\rm track}$ obtained from simulation, with the full size of the correction
taken as a systematic uncertainty, $\pm 0.010$, which is assumed to be fully correlated between bins of the 
differential distributions.
 
\begin{figure}[tb]
  \begin{center}
    \includegraphics[scale=0.5]{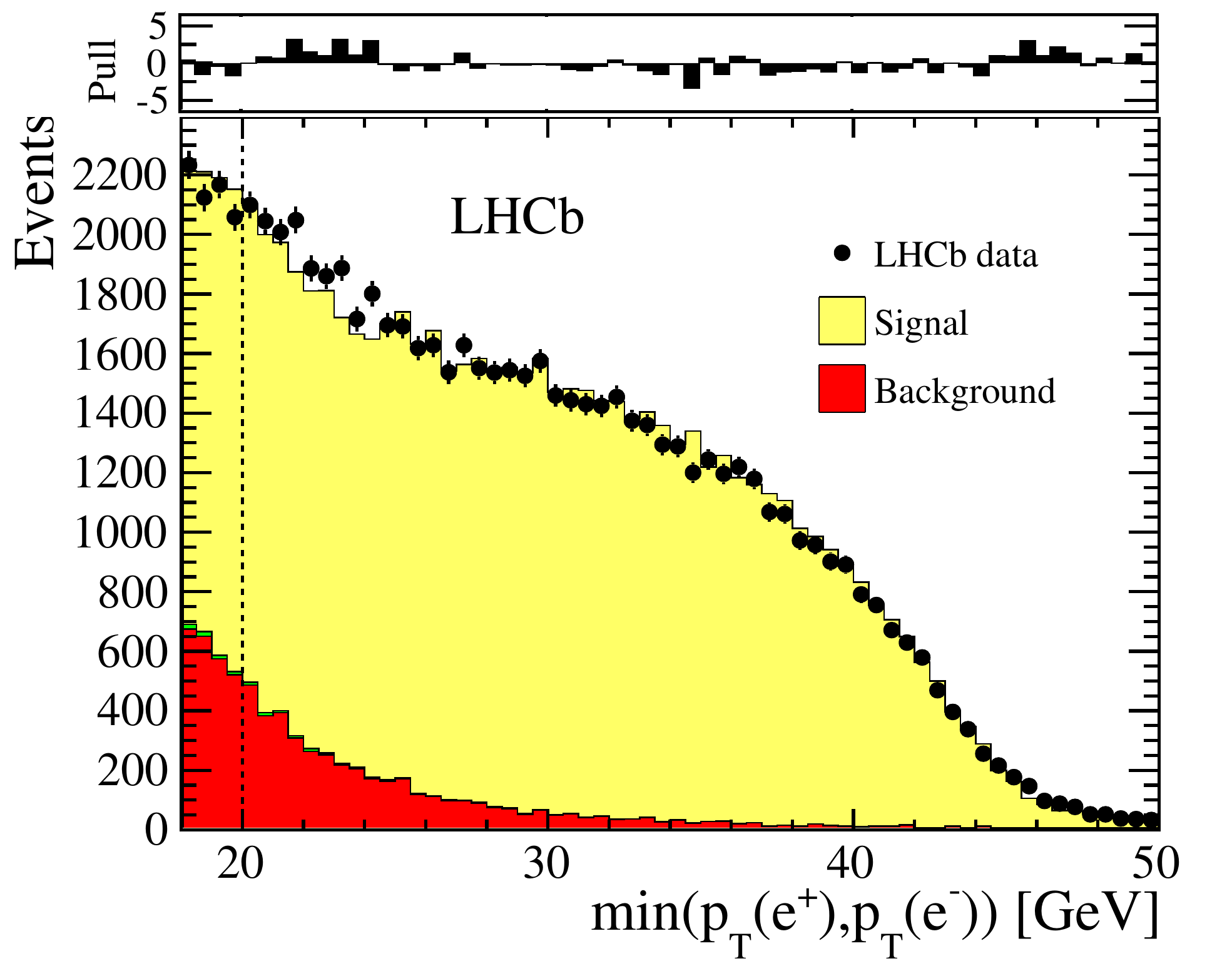}
    \vspace*{-1.0cm}
  \end{center}
  \caption{Comparison between data and simulation for the distribution of 
$\min(\pt(\ep),\pt(\en))$, used in the assessment of uncertainties in  $\epsilon_{\rm kin}$.
The data are shown as points with error bars, the background obtained from same-sign data is shown in red (dark shading), 
to which the expectation from \Z\to\epem\ simulation is added in yellow (light shading).  
The simulated distribution is normalised to the background-subtracted data.  
The $\tautau$ background is also included (green), though barely visible.
The dashed line indicates the threshold applied in the event selection.  The small plot at the top shows the pulls
(\ie\ deviations divided by statistical uncertainties) between the data and the expectation.}
  \label{fig:pTmin}
\end{figure}

The kinematic efficiency is also evaluated from simulation. 
Accurate simulation of the detector material is necessary in order to model correctly 
energy losses through bremsstrahlung, and any inaccuracy would lead to a scaling of
the measured momenta. 
This is tested by  examining the modelling of the \pt\ distributions by simulation, 
particularly in the neighbourhood of the 20\gev threshold. Figure~\ref{fig:pTmin} shows the distribution 
of  $\min(\pt(\ep),\pt(\en))$ for data compared with simulation.  
In order to quantify the uncertainty in $\epsilon_{\rm kin}$, the \pt\ values in data are scaled by a global
factor $\alpha$ to represent the effect of an uncertainty in the detector material. 
The $\chi^2$ between data and simulation is examined as $\alpha$ is varied.
The resulting uncertainty in $\alpha$ translates into a relative uncertainty in $\epsilon_{\rm kin}$ 
of 0.6\% (or 1.2\% for $y_{\Z}>3.75$), which is taken to be a systematic uncertainty fully correlated 
between bins. 

The contribution to $\epsilon_{\rm PID}$ resulting from the calorimeter acceptance is purely geometrical 
and is assumed to be modelled reliably in simulation.   To assess the reliability of simulation for 
the calorimeter energy requirements, events are selected in which one electron 
is tagged using the standard criteria, while a second probe track is found that satisfies all
the requirements apart from that on the energy recorded in one of the calorimeters.  
By examining the distributions of calorimeter energy in the neighbourhood of the threshold applied, 
an estimate of any correction needed and its uncertainty is made.  The test is repeated for each 
part of the calorimeter in turn.
As a result of these studies, a systematic uncertainty 
of 0.7\% on  $\epsilon_{\rm PID}$ is assigned, independent of $y_{\Z}$ and $\phi^*$, 
and treated as fully correlated between bins.
  
The statistical uncertainty in the determination of $\epsilon_{\rm GEC}$  is 
taken as part of the systematic uncertainties. 
The uncertainty in the $10\pm5$ $N_{\rm SPD}$ offset leads to an additional systematic uncertainty of 0.4\% overall,
while an uncertainty of 0.2\% is assigned based on comparing various extrapolation techniques.
The value of  $\epsilon_{\mathrm{GEC}}$ is determined as a function of $y_{\Z}$ and $\phi^*$.
The statistical uncertainty on the determination of $\epsilon_{\mathrm{trig}}$ is 
treated as a systematic uncertainty.  

The principal background to the selected $\Z\to\epem$ sample is expected to arise from failures 
of particle identification, typically where one or two high-\pt\ hadrons interact early and
exhibit a shower profile in the calorimeters similar to electrons.  Such backgrounds are 
addressed by the subtraction of the same-sign $\Z\to\epm\epm$ candidates.  
To check whether this procedure is reliable, event samples are studied in which one electron is tagged 
using the standard requirements, while the second electron satisfies the same criteria except 
that the requirement on HCAL energy is not satisfied, suggesting that the probe is likely to be a hadron.  The numbers 
of same-sign and opposite-sign pairs satisfying these criteria agree 
within 5.5\%.  Treating this as an uncertainty on the background corresponds to an uncertainty of 0.4\% in the 
signal yield, which is taken as the systematic uncertainty on the cross-section.   
There is no significant variation with 
$y_{\Z}$ or $\phi^*$, so the uncertainty is taken to be the same in all bins and fully correlated between bins. 

Physics backgrounds that could give correlated pairs of genuine electron and positron are not necessarily removed 
by the same-sign subtraction.   Production of heavy quark pairs, \ccbar\ or \bbbar, followed by semileptonic 
decay could mimic the signal.  This contribution is expected to be small, 
and is found to be negligible using studies of the distribution of the vertex-fit \chisq
for the candidates in data and simulation.  
The decay $\Z\to\tautau$ provides a background if both $\tau$-leptons decay to electrons.
After the selection, the background from this source is estimated
to be 0.15\% of the signal using simulated samples, and the 
prediction is subtracted in each bin of $y_{\Z}$ and $\phi^*$ as indicated in
Eq.~(\ref{equ:Corr}) with 
its statistical uncertainty included in the statistical error. 
The background arising from production of pairs of gauge bosons, such as $\Wp\Wm$ or $\Wpm\Z$, or of \ttbar pairs,
is neglected, being well below the 0.1\% level, based on expectations from simulation.


\section{Results}
\label{sec:Results}

The measured differential cross-sections as functions of $y_{\Z}$ and $\phi^*$ 
are tabulated in Tables~\ref{tab:YZ8TeV}~and~\ref{tab:Phistar8TeV}.
The uncertainties in the bins of these distributions are significantly 
correlated because the luminosity uncertainty and some of the systematic 
uncertainties are assumed to be common between bins.  The correlation 
matrices between bins of the distributions are presented 
in Tables~\ref{tab:yzcorrel} and~\ref{tab:phistarcorrel} of the Appendix.

The results are given as Born-level cross-sections, which do not include the effects of final-state radiation.  
Tables~\ref{tab:YZ8TeV}~and~\ref{tab:Phistar8TeV} also include the  factors $f_{\rm FSR}$ 
that permit the measurements to be converted to the particle level after FSR.  
These are determined using the true momenta of the
electrons before and after the generation of FSR in the simulation. 

\begin{table}[htb]
  \caption{Differential cross-section for $\Z\to\epem$ as a function of \Z-boson rapidity.
The first error is statistical, the second the uncorrelated experimental systematic, 
the third the correlated experimental systematic and the last error
is the uncertainty in luminosity.  The cross-sections are at the Born level, 
\ie\ before FSR.
The rightmost column gives values of the additional factor, 
$f_{\rm FSR}$, by which the results should be multiplied in order to give the cross-sections after FSR.  }
\begin{center}\begin{tabular}{r@{--}lcr@{$\,\pm\,$}c@{$\,\pm\,$}c@{$\,\pm\,$}c@{$\,\pm\,$}ccc@{$\,\pm\,$}c}
\hline
\multicolumn{2}{c}{$y_{\Z}$} & \multicolumn{7}{c}{${\rm d}\sigma / {\rm d} y_{\Z}$ [pb]} 
& \multicolumn{2}{c}{$f_{\rm FSR}$}\\ 
\hline 
2.000 & 2.125  &&  8.27 & 0.37 & 0.21 & 0.14 & 0.10 && 0.953 & 0.003 \\ 
2.125 & 2.250  && 26.17 & 0.61 & 0.32 & 0.43 & 0.32 && 0.955 & 0.002 \\ 
2.250 & 2.375  && 40.29 & 0.72 & 0.36 & 0.62 & 0.49 && 0.959 & 0.001 \\ 
2.375 & 2.500  && 52.16 & 0.80 & 0.39 & 0.81 & 0.64 && 0.960 & 0.001 \\ 
2.500 & 2.625  && 61.92 & 0.86 & 0.40 & 1.01 & 0.77 && 0.958 & 0.001 \\ 
2.625 & 2.750  && 72.32 & 0.93 & 0.45 & 1.10 & 0.88 && 0.958 & 0.001 \\ 
2.750 & 2.875  && 76.29 & 0.98 & 0.47 & 1.16 & 0.93 && 0.956 & 0.001 \\ 
2.875 & 3.000  && 77.67 & 0.99 & 0.48 & 1.18 & 0.95 && 0.952 & 0.001 \\ 
3.000 & 3.125  && 77.72 & 1.03 & 0.51 & 1.18 & 0.95 && 0.952 & 0.001 \\ 
3.125 & 3.250  && 69.58 & 1.02 & 0.50 & 1.06 & 0.85 && 0.949 & 0.001 \\ 
3.250 & 3.375  && 62.03 & 1.01 & 0.51 & 0.96 & 0.76 && 0.950 & 0.001 \\ 
3.375 & 3.500  && 46.26 & 0.92 & 0.46 & 0.71 & 0.56 && 0.949 & 0.001 \\ 
3.500 & 3.625  && 33.49 & 0.84 & 0.41 & 0.53 & 0.41 && 0.947 & 0.002 \\ 
3.625 & 3.750  && 22.81 & 0.74 & 0.37 & 0.36 & 0.28 && 0.951 & 0.002 \\ 
3.750 & 3.875  && 13.56 & 0.64 & 0.33 & 0.28 & 0.17 && 0.946 & 0.002 \\ 
3.875 & 4.000  &&  6.28 & 0.57 & 0.28 & 0.13 & 0.08 && 0.939 & 0.004 \\ 
4.000 & 4.250  &&  1.85 & 0.33 & 0.16 & 0.04 & 0.02 && 0.928 & 0.005\\
\hline
\end{tabular}\end{center}
\label{tab:YZ8TeV}
\end{table}

\begin{table}[htb]
  \caption{Differential cross-section for $\Z\to\epem$ as a function of $\phi^*$.
The first error is statistical, the second the uncorrelated experimental systematic, 
the third the correlated experimental systematic and the last error
is the uncertainty in luminosity. The cross-sections are at the Born level, 
\ie\ before FSR.
The rightmost column gives values of the additional factor, 
$f_{\rm FSR}$, by which the results should be multiplied in order to give the cross-sections after FSR. }
\begin{center}\begin{tabular}{r@{--}lr@{$\,\pm\,$}c@{$\,\pm\,$}c@{$\,\pm\,$}c@{$\,\pm\,$}cc@{$\,\pm\,$}c}
\hline
\multicolumn{2}{c}{$\phi^*$} & \multicolumn{5}{c}{${\rm d}\sigma / {\rm d} \phi^*$ [pb]}
& \multicolumn{2}{c}{$f_{\rm FSR}$}\\  
\hline 
0.00 & 0.01  &   996 & 13   & 7    & 15   & 12 & 0.954 & 0.001 \\ 
0.01 & 0.02  &   933 & 13   & 7    & 14   & 11 & 0.955 & 0.001 \\ 
0.02 & 0.03  &   851 & 12   & 6    & 13   & 10 & 0.954 & 0.001 \\ 
0.03 & 0.05  &   664 &  8   & 4    & 10    &  8 & 0.954 & 0.001 \\ 
0.05 & 0.07  &   505 &  7   & 3    & 7    &  6 & 0.953 & 0.001 \\ 
0.07 & 0.10  &   346 &  5   & 2    & 5    &  4 & 0.952 & 0.001 \\ 
0.10 & 0.15  & 221.5 & 2.9  & 1.4  & 3.4  & 2.7 & 0.953 & 0.001 \\ 
0.15 & 0.20  & 126.9 & 2.2  & 1.1  & 2.0  & 1.6 & 0.952 & 0.001 \\ 
0.20 & 0.30  &  65.8 & 1.1  & 0.5  & 1.0  & 0.8 & 0.949 & 0.001 \\ 
0.30 & 0.40  &  32.2 & 0.8  & 0.4  & 0.5  & 0.4 & 0.951 & 0.002 \\ 
0.40 & 0.60  & 13.86 & 0.36 & 0.17 & 0.22 & 0.17 & 0.951 & 0.002 \\ 
0.60 & 0.80  &  5.63 & 0.23 & 0.11 & 0.09 & 0.07 & 0.955 & 0.003 \\ 
0.80 & 1.20  &  1.64 & 0.09 & 0.04 & 0.03 & 0.02 & 0.957 & 0.003 \\ 
1.20 & 2.00  &  0.334 & 0.026 & 0.013 & 0.006 & 0.004 & 0.957 & 0.005 \\ 
2.00 & 4.00  &  0.031 & 0.006 & 0.002 & 0.001 & 0.001 & 0.966 & 0.007  \\
\hline
\end{tabular}\end{center}
\label{tab:Phistar8TeV}
\end{table}

The overall cross-section, obtained by integration of the rapidity distribution, is
$$ \sigma(\proton\proton\to\Z\to\epem)=93.81\pm0.41({\rm stat})\pm1.48({\rm syst})\pm1.14({\rm lumi})\;{\rm pb}\,,$$
where the first uncertainty is statistical, the second includes all experimental systematic effects apart from the 
contribution from the luminosity, which forms the third uncertainty.
When combining the experimental systematic uncertainties, 
those associated with $\epsilon_{\rm trig}$, and those parts of $\epsilon_{\rm track}$,
 $\epsilon_{\rm kin}$ and $\epsilon_{\rm PID}$ arising
from the size of the Monte Carlo sample, 
are treated as uncorrelated between bins and therefore combined quadratically; 
other contributions are treated as fully correlated, as is the 
luminosity uncertainty, and combined linearly.

\begin{figure}[htb]
  \begin{center}
    \includegraphics[scale=0.65]{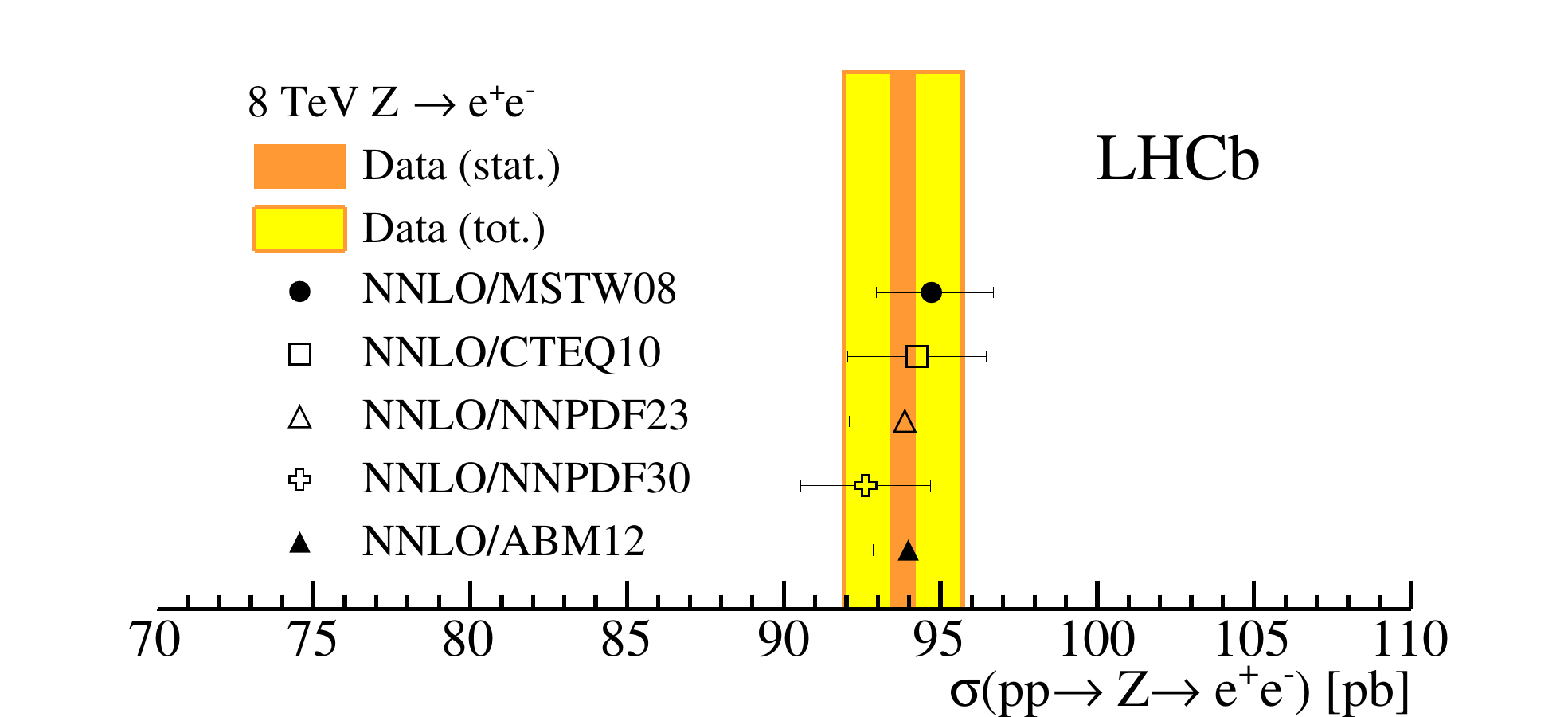}
  \end{center}
  \caption{Measured cross-section for $\Z\to\epem$ shown as the shaded band, with the inner
(orange) band indicating the statistical error and the outer (yellow) band the total uncertainty.  
For comparison, the NNLO predictions of \fewz\ are shown using five different sets of PDFs.
  The uncertainties on these predictions include the PDF uncertainties and the
variation of the factorisation and normalisation scales, as well as the errors arising from numerical integration.}
  \label{fig:ZeeXS}
\end{figure}

The measured cross-section is compared in Fig.~\ref{fig:ZeeXS} with next-to-next-to-leading order (NNLO; $\order(\as^2)$) QCD predictions, 
based on \fewz version 3.1.b2~\cite{Li:2012wna} using five 
different PDF sets, MSTW08~\cite{Martin:2009iq}, CTEQ10~\cite{Lai:2010vv}, ABM12~\cite{Alekhin:2013nda},
NNPDF23~\cite{Ball:2012cx} and NNPDF30~\cite{Ball:2014uwa}.  
Of these, MSTW08 and CTEQ10 predate the start of LHC data-taking, while ABM12 and NNPDF 
have included some LHC measurements in their analyses. 
The uncertainties in the predictions include the effect of varying the renormalisation 
and factorisation scales by factors of two around the nominal values (set to the \Z\ mass), 
combined in quadrature with the PDF uncertainties evaluated at 68\% confidence level. 
All predictions are in good agreement with the data. 

\begin{figure}[htb]
  \begin{center}
    \includegraphics[scale=0.375]{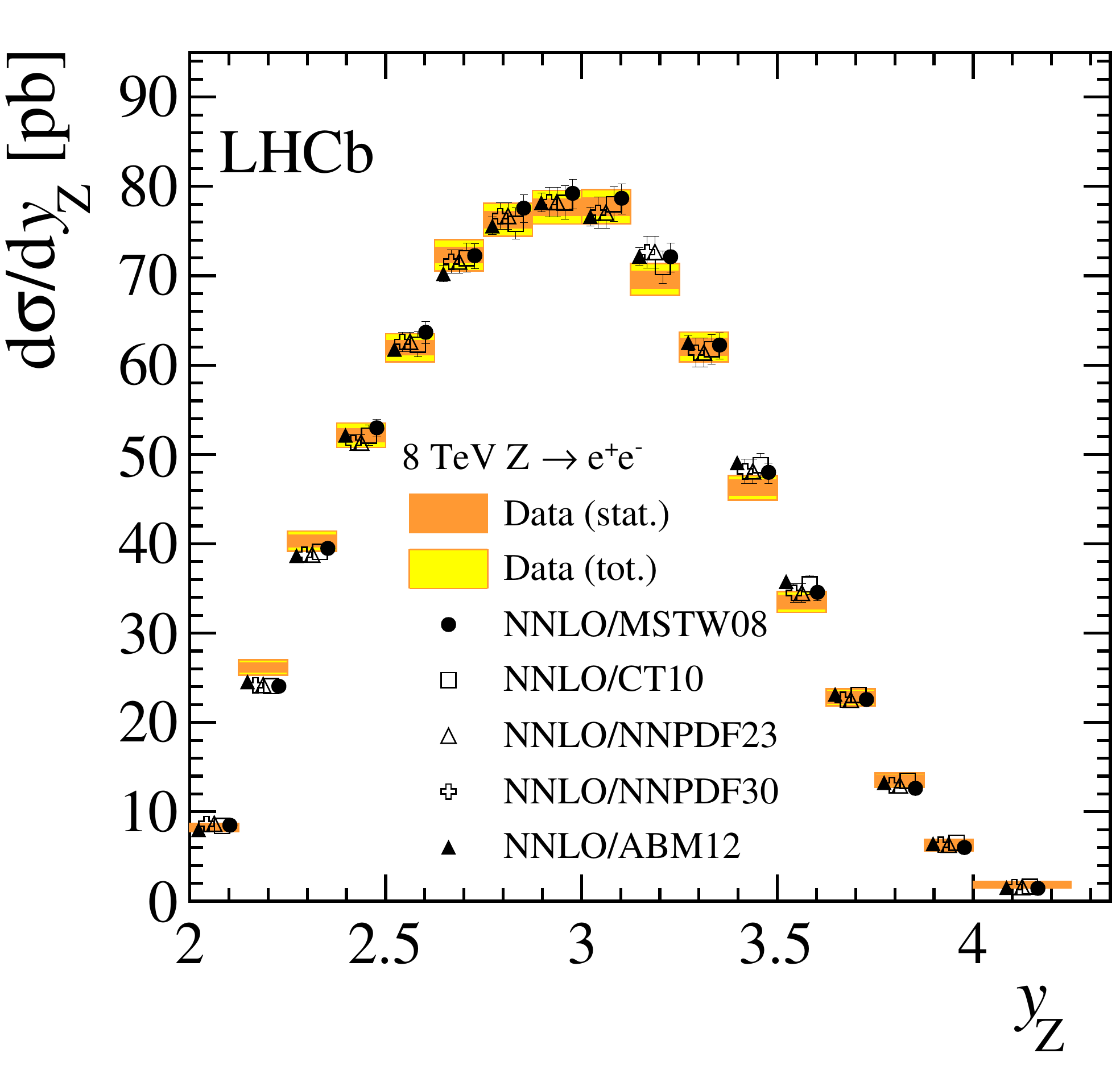}
    \includegraphics[scale=0.375]{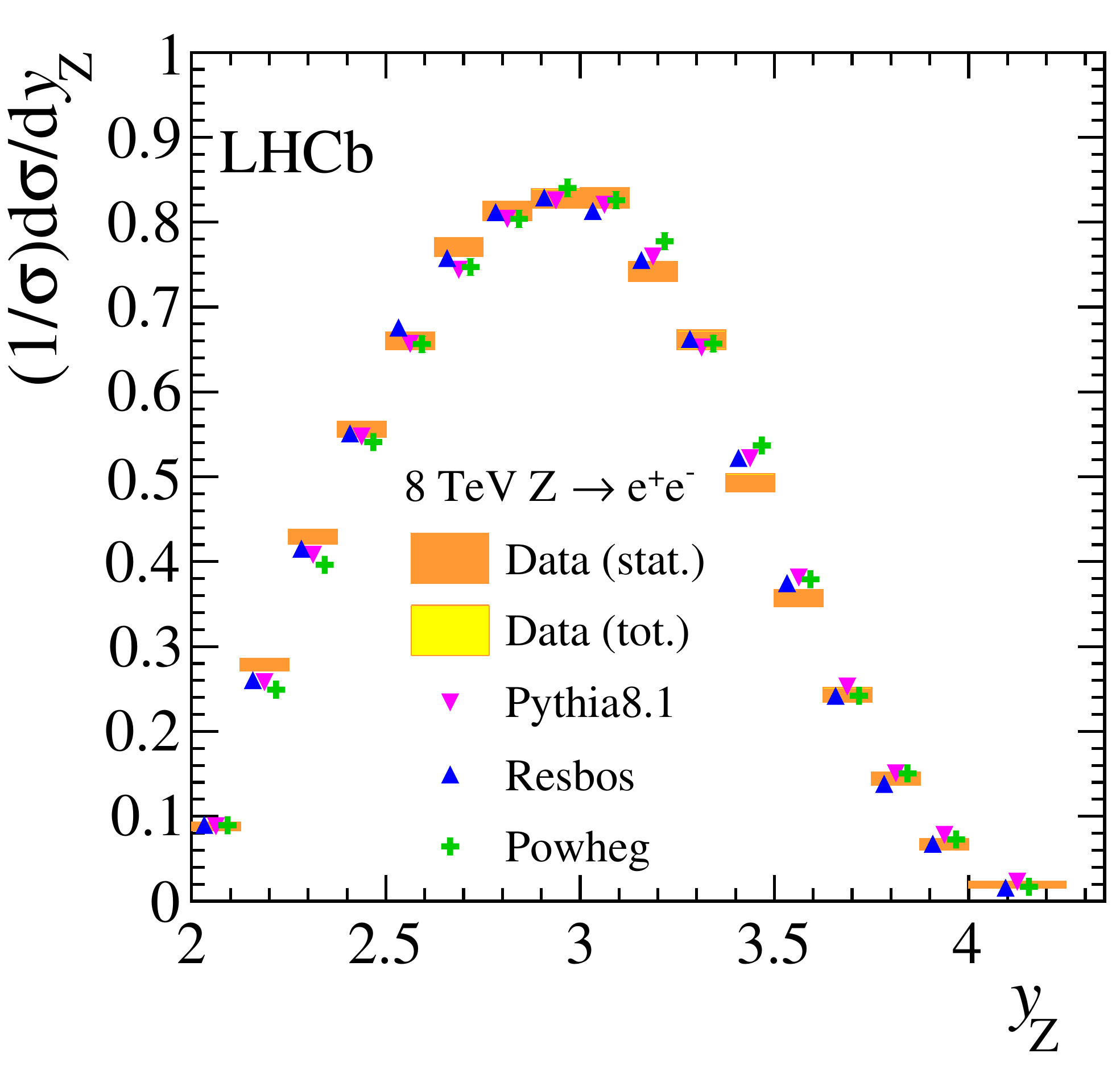}
    \vspace*{-1.0cm}
  \end{center}
  \caption{(left) Differential cross-section ${\rm d}\sigma/{\rm d} y_{\Z}$ and (right)  
normalised differential cross-section $(1/\sigma){\rm d}\sigma/{\rm d} y_{\Z}$ 
as a function of $y_{\Z}$.
The measured  data are shown as the shaded bands, with the inner
(orange) bands indicating the statistical error and the outer (yellow) bands the total uncertainty.  
For comparison, the NNLO predictions of \fewz\ using five 
different sets of PDFs are shown on the left-hand plot. 
The same data are compared with leading log calculations in the right-hand plot.
To aid clarity, small horizontal displacements are applied to some of the predictions.}
  \label{fig:YZ8TeV}
\end{figure}

The differential distribution with respect to    
$y_{\Z}$ is presented in Fig.~\ref{fig:YZ8TeV} (left) and compared with the NNLO calculations based on \fewz, 
all of which are compatible with the integrated cross-section, and model the rapidity distribution as well.
In comparing the shapes of the differential cross-sections with theoretical predictions it can be beneficial 
to normalise them to the integrated cross-section in the acceptance, since most of the correlated systematic
uncertainties in the data cancel.  This is useful when comparing with models based on LO or NLO 
calculations, which may predict the integrated cross-section well.
The normalised differential distribution with respect to   
$y_{\Z}$ is presented in Fig.~\ref{fig:YZ8TeV} (right) and compared 
with calculations that partially take account of higher-order effects.  
A QCD calculation that takes multiple soft gluon emissions 
into account through resummation is provided by \resbos~\cite{Ladinsky:1993zn,*Balazs:1997xd,*Landry
:2002ix}.\footnote{The P branch of \resbos\ is used with 
grids for LHC at $\sqrt{s}=8\tev$ based on CTEQ6.6.}
Alternatively, \powheg~\cite{Alioli:2008gx,*Alioli:2010qp} 
provides a framework whereby a NLO ($\order(\as)$) calculation can be interfaced to a 
parton shower model such as \pythia, which can approximate higher-order effects.  The parton shower model of 
\pythia~8.1~\cite{Sjostrand:2006za,*Sjostrand:2007gs} is also compared with the data.  
All approaches reproduce the main features of the rapidity distribution.

\begin{figure}[htb]
  \begin{center}
    \includegraphics[scale=0.375]{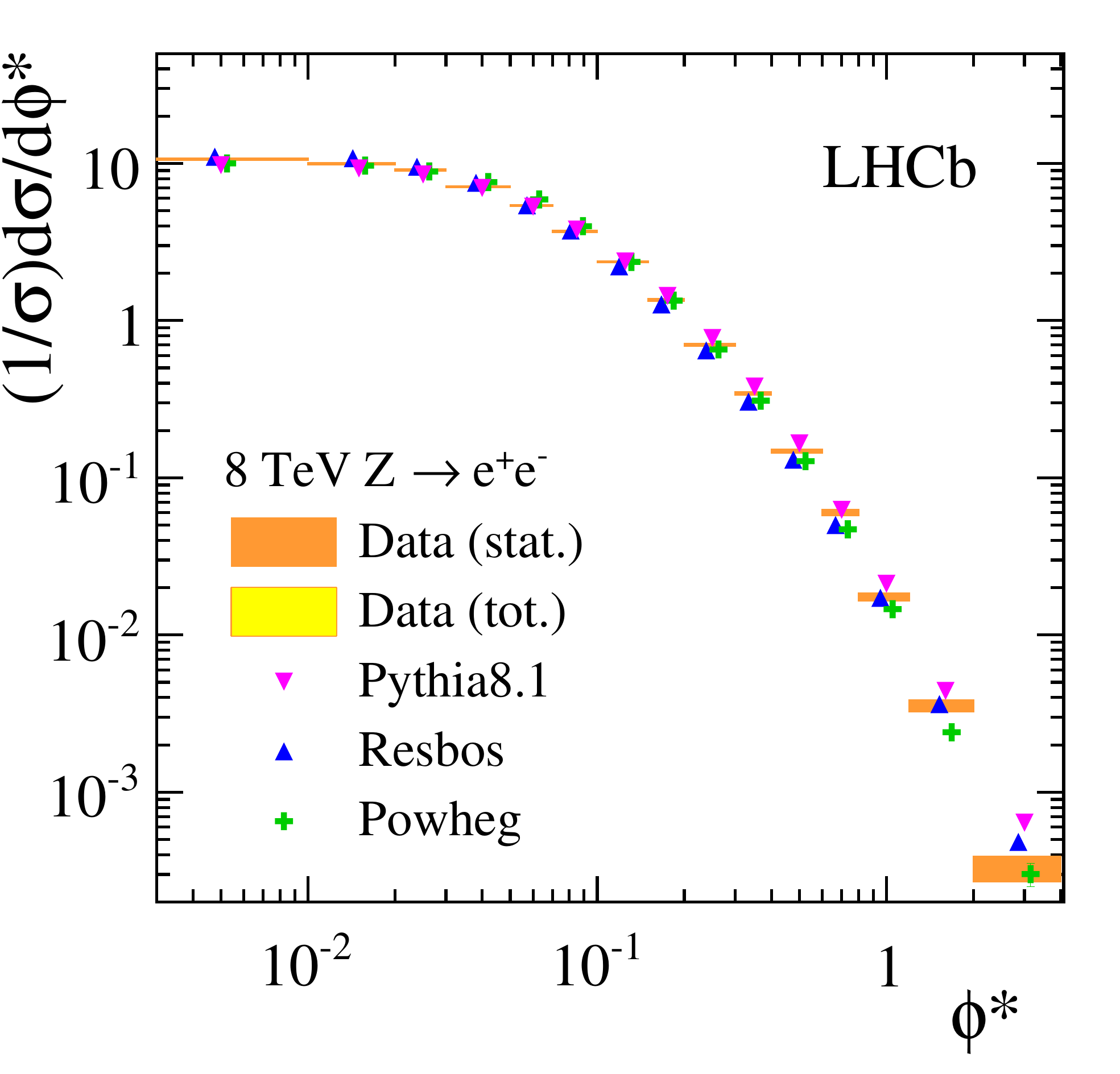}
    \includegraphics[scale=0.375]{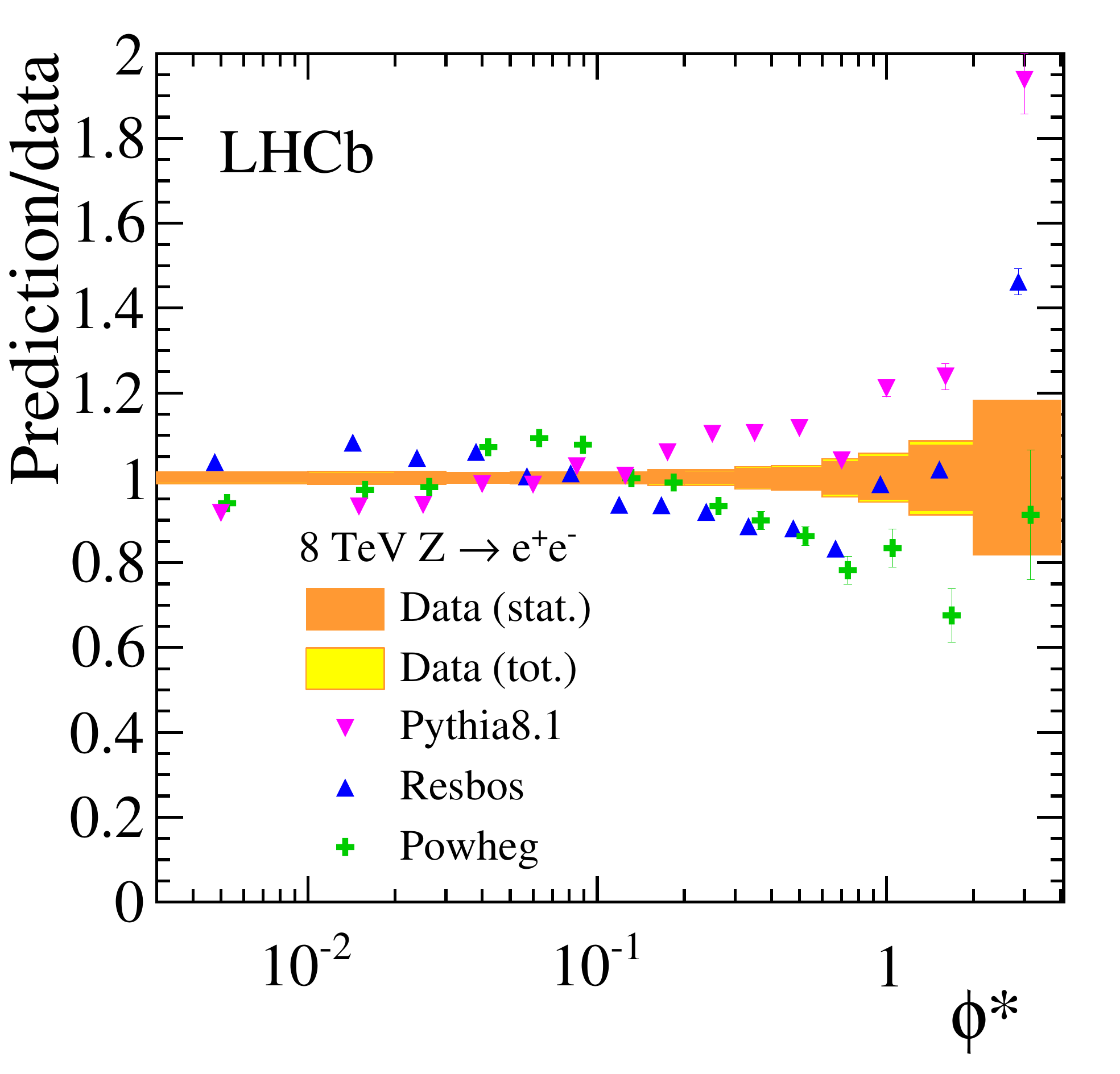}
    \vspace*{-1.0cm}
  \end{center}
  \caption{(left) Normalised  differential cross-section $(1/\sigma){\rm d}\sigma/{\rm d}\phi^*$
 as a function of $\phi^*$.
The measured  data are shown as the shaded bands, with the inner
(orange) bands indicating the statistical error and the outer (yellow) bands the total uncertainty.  
For comparison, the predictions of the leading-log calculations described in the text are shown.
(right) The same data and predictions normalised to the measurement in data, so that
the measurements are shown as the shaded bands at unity. 
To aid clarity, small horizontal displacements are applied to some of the predictions.}
  \label{fig:Phistar8TeV}
\end{figure}

Studies at 7~TeV~\cite{LHCb-PAPER-2012-036}  showed that the NNLO calculations 
based on \fewz fail to model the $\phi^*$ distribution. It is expected that
the $\phi^*$ distribution, like that of \pt, is significantly affected by multiple soft gluon emissions, 
which are not sufficiently accounted for in fixed-order calculations. 
The present data exhibit the same behaviour, and this comparison is not shown. 
The normalised distribution with respect to $\phi^*$ is therefore presented in Fig.~\ref{fig:Phistar8TeV} (left) 
and compared with the 
\resbos, \powheg\ and \pythia~8.1 calculations.
These all model the data reasonably well, especially at lower $\phi^*$ where differences are typically 
up to the 10\% level, while larger discrepancies are seen for $\phi^*>1$. 
To show this more clearly, the ratios between the calculations that include higher orders and the data 
for the $\phi^*$ distribution are also shown in Fig.~\ref{fig:Phistar8TeV} (right).  The data tend to fall 
between the different models, indicating no clear preference for any of them.


\newpage
\section{Summary}
\label{sec:Summary}

A measurement of the cross-section for $\Z$-boson production in the forward region of \proton\proton collisions
at 8\tev centre-of-mass energy is presented.  The measurement, 
using an integrated luminosity of 2.0~\invfb recorded using the \lhcb\ detector, 
is based on the $\Z\to\epem$ decay.  
The acceptance
is defined by the requirements $2.0<\eta<4.5$ and $\pt>20$\gev for the leptons while 
their invariant mass is required to lie in the range
60--120\gev.  The cross-section is determined to be
$$ \sigma(\proton\proton\to\Z\to\epem)=93.81\pm0.41({\rm stat})\pm1.48({\rm syst})\pm1.14({\rm lumi})\;{\rm pb}\,,$$
where the first uncertainty is statistical and the second reflects all systematic uncertainties apart from that associated with
the luminosity, which is given as the third uncertainty.  Differential cross-sections are also presented as functions 
of the Z-boson rapidity, 
and the angular variable $\phi^*$. 
The rapidity distribution is well modelled by NNLO calculations, and is compared with several 
recent sets of parton distribution functions.  A reasonable description of the
$\phi^*$ distribution requires the use of calculations that implement approximations of higher orders,
 either through resummation or using parton shower techniques.

\clearpage







\section*{Acknowledgements}


\noindent We express our gratitude to our colleagues in the CERN
accelerator departments for the excellent performance of the LHC. We
thank the technical and administrative staff at the LHCb
institutes. We acknowledge support from CERN and from the national
agencies: CAPES, CNPq, FAPERJ and FINEP (Brazil); NSFC (China);
CNRS/IN2P3 (France); BMBF, DFG, HGF and MPG (Germany); INFN (Italy); 
FOM and NWO (The Netherlands); MNiSW and NCN (Poland); MEN/IFA (Romania); 
MinES and FANO (Russia); MinECo (Spain); SNSF and SER (Switzerland); 
NASU (Ukraine); STFC (United Kingdom); NSF (USA).
The Tier1 computing centres are supported by IN2P3 (France), KIT and BMBF 
(Germany), INFN (Italy), NWO and SURF (The Netherlands), PIC (Spain), GridPP 
(United Kingdom).
We are indebted to the communities behind the multiple open 
source software packages on which we depend. We are also thankful for the 
computing resources and the access to software R\&D tools provided by Yandex LLC (Russia).
Individual groups or members have received support from 
EPLANET, Marie Sk\l{}odowska-Curie Actions and ERC (European Union), 
Conseil g\'{e}n\'{e}ral de Haute-Savoie, Labex ENIGMASS and OCEVU, 
R\'{e}gion Auvergne (France), RFBR (Russia), XuntaGal and GENCAT (Spain), Royal Society and Royal
Commission for the Exhibition of 1851 (United Kingdom).


\clearpage

{\noindent\bf\Large Appendix}
\section*{Correlation Matrices}
\label{sec:CorrMat}
 
\begin{table}[tbh]
\caption{Correlation matrix between bins of $y_{\rm Z}$.
The bin numbering follows the same sequence as Table~\ref{tab:YZ8TeV}.}
\begin{center}
\scalebox{0.75}{
\begin{tabular}{c|ccccccccccccccccc}
\hline
 Bin & \\
 index &	 1		&	 2		&	 3		&	 4		&	 5		&	 6		&	 7		&	 8		&	 9		&	 10		&	 11		&	 12		&	 13		&	 14		&	 15		&	 16		&	 17	\\	\hline
 1&	1.00		&			&			&			&			&			&			&			&			&			&			&			&			&			&			&			&			\\
 2&	0.23		&	1.00		&			&			&			&			&			&			&			&			&			&			&			&			&			&			&			\\
 3&	0.26		&	0.43		&	1.00		&			&			&			&			&			&			&			&			&			&			&			&			&			&			\\
 4&	0.22		&	0.36		&	0.41		&	1.00		&			&			&			&			&			&			&			&			&			&			&			&			&			\\
 5&	0.30		&	0.49		&	0.56		&	0.47		&	1.00		&			&			&			&			&			&			&			&			&			&			&			&			\\
 6&	0.30		&	0.50		&	0.56		&	0.47		&	0.64		&	1.00		&			&			&			&			&			&			&			&			&			&			&			\\
 7&	0.30		&	0.50		&	0.57		&	0.47		&	0.65		&	0.65		&	1.00		&			&			&			&			&			&			&			&			&			&			\\
 8&	0.30		&	0.50		&	0.57		&	0.47		&	0.65		&	0.65		&	0.65		&	1.00		&			&			&			&			&			&			&			&			&			\\
 9&	0.30		&	0.49		&	0.56		&	0.47		&	0.64		&	0.64		&	0.65		&	0.64		&	1.00		&			&			&			&			&			&			&			&			\\
 10&	0.28		&	0.47		&	0.54		&	0.45		&	0.61		&	0.62		&	0.62		&	0.62		&	0.61		&	1.00		&			&			&			&			&			&			&			\\
 11&	0.27		&	0.45		&	0.51		&	0.43		&	0.58		&	0.59		&	0.59		&	0.59		&	0.58		&	0.56		&	1.00		&			&			&			&			&			&			\\
 12&	0.25		&	0.41		&	0.46		&	0.39		&	0.53		&	0.53		&	0.54		&	0.54		&	0.53		&	0.51		&	0.49		&	1.00		&			&			&			&			&			\\
 13&	0.22		&	0.36		&	0.41		&	0.34		&	0.47		&	0.47		&	0.47		&	0.47		&	0.47		&	0.45		&	0.43		&	0.39		&	1.00		&			&			&			&			\\
 14&	0.18		&	0.30		&	0.34		&	0.28		&	0.39		&	0.39		&	0.39		&	0.39		&	0.39		&	0.37		&	0.35		&	0.32		&	0.28		&	1.00		&			&			&			\\
 15&	0.15		&	0.26		&	0.29		&	0.24		&	0.33		&	0.33		&	0.33		&	0.33		&	0.33		&	0.32		&	0.30		&	0.27		&	0.24		&	0.20		&	1.00		&			&			\\
 16&	0.09		&	0.14		&	0.16		&	0.14		&	0.18		&	0.19		&	0.19		&	0.19		&	0.18		&	0.18		&	0.17		&	0.15		&	0.13		&	0.11		&	0.10		&	1.00		&			\\
 17&	0.04		&	0.07		&	0.08		&	0.07		&	0.10		&	0.10		&	0.10		&	0.10		&	0.10		&	0.09		&	0.09		&	0.08		&	0.07		&	0.06		&	0.05		&	0.03		&	1.00		\\
\hline
\end{tabular}
}
\end{center}
\label{tab:yzcorrel}

\end{table}

\begin{table}[tbh]
\caption{Correlation matrix between bins of $\phi^*$.
The bin numbering follows the same sequence as Table~\ref{tab:Phistar8TeV}.}
\begin{center}
\scalebox{0.8}{
\begin{tabular}{c|ccccccccccccccc}
\hline
Bin index &	 1		&	 2		&	 3		&	 4		&	 5		&	 6		&	 7		&	 8		&	 9		&	 10		&	 11		&	 12		&	 13		&	 14		&	 15							\\\hline
 1&	1.00		&			&			&			&			&			&			&			&			&			&			&			&			&			&									\\
 2&	0.63		&	1.00		&			&			&			&			&			&			&			&			&			&			&			&			&									\\
 3&	0.62		&	0.61		&	1.00		&			&			&			&			&			&			&			&			&			&			&			&									\\
 4&	0.82		&	0.80		&	0.79		&	1.00		&			&			&			&			&			&			&			&			&			&			&									\\
 5&	0.64		&	0.63		&	0.62		&	0.82		&	1.00		&			&			&			&			&			&			&			&			&			&									\\
 6&	0.64		&	0.63		&	0.62		&	0.82		&	0.64		&	1.00		&			&			&			&			&			&			&			&			&									\\
 7&	0.65		&	0.64		&	0.63		&	0.83		&	0.65		&	0.65		&	1.00		&			&			&			&			&			&			&			&									\\
 8&	0.58		&	0.57		&	0.56		&	0.74		&	0.58		&	0.58		&	0.59		&	1.00		&			&			&			&			&			&			&									\\
 9&	0.58		&	0.58		&	0.57		&	0.75		&	0.59		&	0.59		&	0.59		&	0.53		&	1.00		&			&			&			&			&			&									\\
 10&	0.48		&	0.48		&	0.47		&	0.62		&	0.49		&	0.49		&	0.49		&	0.44		&	0.44		&	1.00		&			&			&			&			&									\\
 11&	0.46		&	0.46		&	0.45		&	0.59		&	0.47		&	0.47		&	0.47		&	0.42		&	0.43		&	0.35		&	1.00		&			&			&			&									\\
 12&	0.35		&	0.34		&	0.34		&	0.44		&	0.35		&	0.35		&	0.35		&	0.31		&	0.32		&	0.26		&	0.25		&	1.00		&			&			&									\\
 13&	0.30		&	0.30		&	0.30		&	0.39		&	0.31		&	0.31		&	0.31		&	0.28		&	0.28		&	0.23		&	0.22		&	0.16		&	1.00		&			&									\\
 14&	0.22		&	0.22		&	0.21		&	0.28		&	0.22		&	0.22		&	0.22		&	0.20		&	0.20		&	0.17		&	0.16		&	0.12		&	0.10		&	1.00		&									\\
 15&	0.10		&	0.10		&	0.10		&	0.13		&	0.10		&	0.10		&	0.10		&	0.09		&	0.09		&	0.08		&	0.07		&	0.06		&	0.05		&	0.04		&	1.00								\\
\hline
\end{tabular}
\label{tab:phistarcorrel}
}
\end{center}
\end{table}

\clearpage

\addcontentsline{toc}{section}{References}
\setboolean{inbibliography}{true}
\bibliographystyle{LHCb}
\bibliography{main,ZeeBib,LHCb-PAPER,LHCb-DP}

\newpage

\centerline{\large\bf LHCb collaboration}
\begin{flushleft}
\small
R.~Aaij$^{41}$, 
B.~Adeva$^{37}$, 
M.~Adinolfi$^{46}$, 
A.~Affolder$^{52}$, 
Z.~Ajaltouni$^{5}$, 
S.~Akar$^{6}$, 
J.~Albrecht$^{9}$, 
F.~Alessio$^{38}$, 
M.~Alexander$^{51}$, 
S.~Ali$^{41}$
% ...

{\footnotesize \it
% ...
$ ^{5}$Clermont Universit\'{e}, Universit\'{e} Blaise Pascal, CNRS/IN2P3, LPC, Clermont-Ferrand, France\\
$ ^{6}$CPPM, Aix-Marseille Universit\'{e}, CNRS/IN2P3, Marseille, France\\
% ...
$ ^{9}$Fakult\"{a}t Physik, Technische Universit\"{a}t Dortmund, Dortmund, Germany\\
% ...
$ ^{37}$Universidad de Santiago de Compostela, Santiago de Compostela, Spain\\
$ ^{38}$European Organization for Nuclear Research (CERN), Geneva, Switzerland\\
% ...
$ ^{41}$Nikhef National Institute for Subatomic Physics, Amsterdam, The Netherlands\\
% ...
$ ^{46}$H.H. Wills Physics Laboratory, University of Bristol, Bristol, United Kingdom\\
% ...
$ ^{51}$School of Physics and Astronomy, University of Glasgow, Glasgow, United Kingdom\\
$ ^{52}$Oliver Lodge Laboratory, University of Liverpool, Liverpool, United Kingdom\\
% ...
}
\end{flushleft}


\end{document}